\documentclass[useAMS,usenatbib]{mn2e}
\usepackage{aas_macros,graphicx,times,multirow,amsmath}
\title[The monitoring of \src]{The discovery, monitoring and environment of \src}
\author[G.L.~Israel et al.] {G.L.~Israel,$^{1}$\thanks{E-mail: gianluca@oa-roma.inaf.it} P.~Esposito,$^{2,3}$ N.~Rea,$^{4,5}$ F.~Coti Zelati,$^{6,4,7}$ A.~Tiengo,$^{8,9,10}$ \newauthor S.~Campana,$^{7}$ S.~Mereghetti,$^{8}$ G.A.~Rodriguez Castillo,$^{1}$ D. G\"otz,$^{11}$ M. Burgay,$^{12}$ \newauthor A. Possenti,$^{12}$ S. Zane,$^{13}$ R. Turolla,$^{14,13}$ 
R. Perna,$^{15}$ G. Cannizzaro,$^{16}$ and J. Pons $^{17}$  
\smallskip\\
$^1 $Osservatorio Astronomico di Roma, INAF, via Frascati 33, I-00040 Monteporzio Catone, Italy\\
$^2$ Istituto di Astrofisica Spaziale e Fisica Cosmica - Milano, INAF, via E. Bassini 15, I-20133 Milano, Italy\\
$^3$ Harvard-Smithsonian Center for Astrophysics, 60 Garden Street, Cambridge, MA 02138, USA\\
$^{4}$ Anton Pannekoek Institute for Astronomy, University of Amsterdam, Postbus 94249,  NL-1090-GE Amsterdam, The Netherlands\\
$^{5}$ Instituto de Ciencias de l'Espacio (ICE, CSIC--IEEC), Carrer de Can Magrans, S/N, 08193, Barcelona, Spain\\
$^{6}$ Universit\`a dell'Insubria, via Valleggio 11, I-22100 Como, Italy\\
$^{7}$ INAF -- Osservatorio Astronomico di Brera, via Bianchi 46, I-23807 Merate (LC), Italy\\
$^{8}$ INAF -- Istituto di Astrofisica Spaziale e Fisica Cosmica, via E. Bassini 15, I-20133 Milano, Italy\\
$^9$ IUSS -- Istituto Universitario di Studi Superiori, piazza della Vittoria 15, I-27100 Pavia, Italy\\
$^{10}$ INFN -- Istituto Nazionale di Fisica Nucleare, Sezione di Pavia, via A. Bassi 6, I-27100 Pavia, Italy\\
$^{11}$ AIM CEA/Irfu/Service d'Astrophysique, Orme des Merisiers, F-91191 Gif-sur-Yvette, France\\
$^{12}$ Osservatorio Astronomico di Cagliari, via della Scienza 5, 09047, Cagliari, Italy \\
$^{13}$ Mullard Space Science Laboratory, University College London, Holmbury St. Mary, Dorking, Surrey RH5 6NT, UK\\
$^{14}$ Dipartimento di Fisica e Astronomia, Universit\`a di Padova, via F. Marzolo 8, I-35131 Padova, Italy\\
$^{15}$ Department of Physics and Astronomy, Stony Brook University, Stony Brook, NY 11794, USA\\
$^{16}$ Universit\`a di Roma "La Sapienza", P.le A. Moro 5, 00185, Roma, Italy \\
$^{17}$ Departament de F\`isica Aplicada, Universitat d' Alacant, Ap. Correus 99, 03080 Alacant, Spain\\
}

\def\LaTeX{L\kern-.36em\raise.3ex\hbox{a}\kern-.15em
    T\kern-.1667em\lower.7ex\hbox{E}\kern-.125emX}
\def\xmm {\emph{XMM-Newton}}
\def\cxo {\emph{Chandra}}
\def\swift {\emph{Swift}}

\def\xte {\emph{RXTE}}

\def\src {SGR\,J1935+2154}

\def\rxs {1RXS\,J1708$-$4009}
\def\xte{XTE\,J1810$-$197}

\def\sgrj{Swift\,J1834$-$0846}

\def\rrata {RRAT\,J1819$-$1458}
\def\flux {\mbox{erg cm$^{-2}$ s$^{-1}$}}
\def\lum {\mbox{erg s$^{-1}$}}
\def\nh {$N_{\rm H}$}
\def\rc {\rm}

\begin{document}

\label{firstpage}
\maketitle
\begin{abstract}
We report on the discovery of a new member of the magnetar class, \src, and on its timing and spectral properties measured by an extensive observational campaign carried out between {\rc July 2014 and March 2015} with \cxo\ and \xmm\ (11 pointings).

We discovered the spin period of \src\ through the detection of coherent pulsations at a period of about 3.24\,s. 
The magnetar is slowing-down at a rate of $\dot{P}$=$1.43(1)\times$10$^{-11}$ s s$^{-1}$ and with a decreasing trend due to a negative   $\ddot{P}$ of $-3.5(7)\times$10$^{-19}$ s s$^{-2}$. This implies a surface dipolar magnetic field strength of $\sim$2.2$\times$10$^{14}$ G, a characteristic age of about 3.6\,kyr and, a spin-down luminosity  L$_{sd}\sim$1.7$\times$10$^{34}$\lum. The source spectrum is well modelled by a blackbody with temperature of about 500\,eV  plus a power-law component with photon index of about 2. The source showed a moderate long-term variability, with a flux decay of about 25\% during the first four months since its discovery, {\rc and a re-brightening of the same amount during the second four months}. 

The X-ray data were also used to study the source environment. In particular, we discovered a diffuse emission extending on spatial scales from about 1$^{\prime\prime}$ up to at least 1$^{\prime}$ around \src\  both in \cxo\  and \xmm\ data. This component is constant in flux (at least within uncertainties) and its spectrum is well modelled by a power-law spectrum  steeper than that of the pulsar. Though a scattering halo origin seems to be more probable we cannot exclude that part, or all, of the diffuse emission is due to a pulsar wind nebula. 
\end{abstract}

\section{Introduction}

Large observational and theoretical efforts have been devoted in the
past years to unveil the nature of a sample of peculiar high-energy
pulsars, namely the Anomalous X-ray Pulsars (AXPs) and the Soft
Gamma-ray Repeaters (SGRs). These objects are believed to be isolated neutron stars and powered 
by their own magnetic energy, stored in a super-strong field, and are collectively 
referred to as magnetars \citep*{duncan92,paczynski92}.   
They share similar
timing properties (spin period $P$ in the 2--12 s range and period derivative
$\dot{P}$ in the 10$^{-13}$--10$^{-11}$ s\,s$^{-1}$, range). Their X-ray
luminosity, typically $L_X$ $\sim$ 10$^{33}$--10$^{35}$ \lum,
generally exceeds the rotational energy loss rate, while  the
temperatures of the thermal component observed in their spectra are
often higher than those predicted by models of non-magnetic cooling neutron stars.
Their (surface dipolar) magnetic fields inferred from the dipolar-loss formula
are generally  of the order of B$\sim$10$^{14}$ -- 10$^{15}$ G.
However, recently low dipole field magnetars have been discovered,
which behave as typical magnetars but with dipolar
magnetic field as low as $6\times10^{12}$\,G, i.e. in the range of normal
radio pulsars \citep{rea10}: these sources possibly store large magnetic
energy in other components of their magnetic field \citep{turolla11,rea13}. 

Sporadically, magnetars emit high energy (up to the MeV range) bursts and
flares which can last from a fraction of a seconds to minutes, releasing
$\sim$10$^{38}\div$10$^{47}$ erg s$^{-1}$, often accompanied
by long-lived (up to years) increases of the persistent X-ray
luminosity (outbursts). These events may be accompanied or triggered
by deformations or fractures of the neutron star crust  and/or
local/global rearrangements of the star magnetic field. The detection of these
energetic events provides the main channel to identify new objects
of this class.

A fundamental question about magnetar  concerns their
evolutionary link to their less magnetic 
siblings, the rotation-powered pulsars. A number of unexpected results,
both from known and newly discovered magnetars,  drastically
changed our understanding of these objects. In 2004, while studying the
emission properties of the bright X-ray transient magnetar \xte,
the source was discovered to be a bright transient radio pulsar, the
first of the class \citep{camilo06}. Today we know that 4 out of
the about 25 known magnetars, are occasionally shining as radio
pulsars in the outburst phase. All the radio ``active" magnetars are
characterized by a quiescent X-ray over spin-down luminosity ratio of
$L_X/L_{sd}<1$ \citep{rea12}.


\begin{table}
\centering \caption{Summary of the {\rc \swift}, \cxo\ and \xmm\ observations used in this work and carried out {between July 2014 and March 2015}.} \label{logs}
\begin{tabular}{@{}llcc}
\hline
Mission~/~Obs.\,ID & Instrument & Date & Exposure \\ 
 & & & (ks) \\ 
\hline
{\rc \swift~/603488000} & {\rc XRT}  & {\rc Jul 5}  & {\rc 3.4}  \\
{\rc \swift~/603488002}  & {\rc XRT}  & {\rc Jul 6}  & {\rc 4.3}  \\
{\rc \swift~/603488004}  & {\rc XRT}  & {\rc Jul 7}  & {\rc 9.3}  \\
{\rc \swift~/603488006}  & {\rc XRT}  & {\rc Jul 8}  & {\rc 3.7}  \\
{\rc \swift~/603488008}  & {\rc XRT}  & {\rc Jul 13}  & {\rc 5.3}  \\
{\rc \swift~/603488009}  & {\rc XRT}  & {\rc Jul 13 } & {\rc 3.0}  \\
\cxo~/~15874 & ACIS-S &  Jul 15 & 10.1  \\ 
{\rc \swift~/603488010}  & {\rc XRT}  & {\rc Jul 16}  & {\rc 7.1 } \\
\cxo~/~15875 & ACIS-S$^a$ & Jul 28 & 75.4 \\ 
\cxo~/~17314 & ACIS-S$^a$  & Aug 31 & 29.2 \\ 
{\em XMM}~/~0722412501 & EPIC & Sep 26 & 19.0 \\ 
{\em XMM}~/~0722412601 & EPIC & Sep 28 & 20.0 \\ 
{\em XMM}~/~0722412701 & EPIC & Oct 04 & 18.0 \\ 
{\em XMM}~/~0722412801 & EPIC & Oct 16 & 9.7 \\ 
{\em XMM}~/~0722412901 & EPIC & Oct 24 & 7.3 \\ 
{\em XMM}~/~0722413001 & EPIC & Oct 27 & 12.6 \\ 
{\em XMM}~/~0748390801 & EPIC & Nov 15 & 10.8 \\ 
{\rc {\em XMM}~/~0764820101} & {\rc EPIC} & {\rc Mar 25 } & {\rc 28.4 }\\ 
\hline
\end{tabular}
\begin{list}{}{}
\item[$^{a}$] Data collected in continuous clocking mode (CC).
\end{list}
\end{table}

Energetic pulsars are known to produce particle outflows, often
resulting in spectacular pulsar wind nebulae (PWNe) of which the Crab
is the most famous example \citep{weisskopf00}. Magnetars are
expected to produce particle outflows as well, either in quiescence or during
outbursts accompanying bright bursts. Given the strong magnetic fields
associated with this class of neutron stars, the idea of a wind nebula 
around a magnetar is thus promising. There has not been yet a
confirmed detection of such a nebula, but some cases of
``magnetically powered'' X-ray nebulae around pulsars with relatively high
magnetic fields have been suggested. A peculiar extended emission 
has been reported around the rotating
radio transient RRAT J1819-1458 \citep{rea09,camero13}, 
with a nominal X-ray efficiency $\eta_X \sim 0.2$, too high to
be only rotationally powered. The authors 
suggested that the occurrence of the nebula might be connected with
the high magnetic field (B = 5$\times$10$^{13}$ G) of the pulsar.
Similarly, \cite{younes12}  reported the discovery of a possible
wind nebula around \sgrj, with an X-ray efficiency $\eta_X \sim 0.7$ 
(but see \citealt{etr13} for a different interpretation in terms of dust scatter). 

\src\ is a newly discovered member of the magnetar family, and was
discovered thanks to the detection of low-Galactic latitude short
bursts by \swift\ on 2014 July 5 \citep{stamatikos14}. Follow-up
observations carried out by \cxo\ on 2014 July 15  and 29  allowed
us to precisely locate the source and  detect its spin period
($P$=3.25\,s; \citealt{israel14}) confirming that \src\ is indeed a
magnetar. The \src\ position is coincident with the center of the 
Galactic supernova remnant (SNR) G57.2+0.8 of undetermined age 
and at a {\rc possible, but uncertain,} distance of 9\,kpc \citep{sun11,pavlovic13}.

In this paper we report on the results of an \xmm\, and \cxo\,
observational campaign covering the first {\rc 8} months of \src 's
outburst. Our observational campaign is ongoing with \xmm, and its long-term results will be reported elsewhere. 
We also report upper limits on the radio emission derived  
from Parkes observations \citep{burgay14}. We first report on the
data analysis, then summarize the results we obtained for the 
parameters, properties and environment of this new magnetar. Finally we discuss
 our findings in the contest of the magnetar scenario.

\section{X--ray Observations}\label{observations}

\subsection{\cxo}\label{acisdata}
\cxo\ observations of \src\ were  carried out  three times during July and August 2014 (see Table~\ref{logs}) in response to the detection of short SGR-like bursts from the source. The first dataset was acquired with the ACIS-S instrument in Faint imaging (Timed Exposure)  and 1/8 subarray mode (time resolution: $\sim$0.44 s), while the subsequent two pointings were obtained with the ACIS-S in Faint timing (Continuous Clocking) mode (time resolution 2.85 ms).

The data were reprocessed with the Chandra Interactive Analysis of Observations software (\textsc{ciao}, version 4.6) using the calibration files available in the \cxo\ \textsc{caldb} 4.6.3 database. The scientific products were extracted following standard procedures,  but adopting extraction regions with different size in order to properly subtract the underlying diffuse component (see Section \ref{diffuse} and Figure \ref{ima_psf}). 
Correspondingly, for the first observation (Faint imaging) we used circular regions of 1.5$^{\prime \prime}$ (and 3.0$^{\prime \prime}$) radius for the source (and diffuse emission) associated to a background annular region  with 1.6$^{\prime \prime}$ and 3.0$^{\prime \prime}$ (10$^{\prime \prime}$, 15$^{\prime \prime}$) for the inner and outer radius, respectively.
Furthermore we used rectangular boxes of 3$^{\prime \prime}$$\times$2$^{\prime \prime}$ (and 4$^{\prime \prime}$$\times$2$^{\prime \prime}$) sides aligned to the CCD readout direction for the remaining two observations in CC mode. For the background we used two rectangular boxes of 1.5$^{\prime \prime}$ $\times$1.5$^{\prime \prime}$ (and 2$^{\prime \prime}$ $\times$2$^{\prime \prime}$) at the sides of the source extraction region. 
For the spectra, the  redistribution matrices and the ancillary response files were created using \textsc{specextract}. For the timing analysis, we applied the Solar system barycentre correction to the photon arrival times with \textsc{axbary}.

\subsection{\xmm}\label{xmmdata}
\xmm\ observations of \src\ were carried out between September and March 2015 (see Table~\ref{logs}) to monitor the source decay and study the source  properties. We used the data collected with the European Photon Imaging Camera  (\textsc{epic}), which consists of two MOS \citep{turner01}  and one pn \citep{struder01}  CCD detectors. The raw data were reprocessed using the \xmm\ Science Analysis Software (\textsc{sas}, version 14.0) and the calibration files in the \textsc{ccf} release of 2015 March. The pn operated in Full Window (time resolution of about 73 ms) while the MOSs were set in Small Window (time resolution of 300 ms), therefore optimized for the timing analysis.
The intervals of flaring background were located by intensity filters (see e.g. \citealt{deluca04}) and excluded from the analysis. Source photons were extracted from circles with radius of 40$^{\prime\prime}$. The pn background was extracted from an annular region with inner and outer radii of 45$^{\prime\prime}$ and 90$^{\prime\prime}$, respectively (also in this case the choice was dictated by the diffuse emission component; Section \ref{diffuse} and Figure \ref{ima_psf}).  Photon arrival times were converted to the Solar system barycenter using the \textsc{sas} task \textsc{barycen} using the source coordinate as inferred from the \cxo\ pointings (see Section\,\ref{where}). The ancillary response files and the spectral redistribution matrices for the spectral analysis were generated with \textsc{arfgen} and \textsc{rmfgen}, respectively. In order to maximize the signal to noise ratio we combined, when needed, the spectra from the available EPIC cameras and averaged the response files using \textsc{epicspeccombine}. In particular, the latter command was routinely applied for the study of the dim diffuse emission.

\subsection{\swift}
{\rc The \swift\ X-Ray Telescope (XRT) uses a front-illuminated CCD detector sensitive to photons 
between 0.2 and 10 keV  \citep{burrows05} . Two readout modes can be used: photon counting (PC) and windowed timing (WT). The PC mode provides images and a 2.5\,s  time resolution; in WT mode only one-dimensional imaging is  preserved with a time resolution of 1.766\,ms. Data were processed with \textsc{xrtpipeline} (version 12), and ﬁltered and  screened with standard criteria, correcting for effective area,  dead columns, etc. Events were extracted from a  20 pixel radius region around the source position. For  spectroscopy we used the spectral redistribution matrices in \textsc{caldb} (20130101, v014 for the PC), while the ancillary response ﬁles  were generated with \textsc{xrtmkarf}.}

\section{Analysis and results}\label{analysis}

\subsection{Position}\label{where}
We used the \cxo\ ACIS-S observation carried out on 2014 July the 15th, the only one in imaging mode, in order to precisely locate \src. Only one bright source was detected in the S7 CCD operating at 1/8 of the nominal field of view. The refined position of the source, calculated with \textsc{wavdetect}, is R.A. = 19$^h$34$^m$55$\fs$5978s, Dec. = +21$^o$53$^{\prime}$47$\farcs$7864 (J2000.0; statistical uncertainty of 0$\farcs$02) with a 90\% confidence level uncertainty radius of 0$\farcs$7. This position is consistent with that of \src\  obtained by \swift: R.A. = 19$^h$34$^m$55$\fs$68s, Dec. = +21$^o$53$^{\prime}$48$\farcs$2, J2000.0, radius of 2$\farcs$3 at 90\% confidence level \citep{cummings14}. Correspondingly, we are confident that the source we detected in the \cxo\ image is indeed the source first detected by \swift\ BAT and later by XRT and responsible for the observed SGR-like bursts.

\subsection{Spatial analysis}\label{diffuse}
\begin{figure}
\centering
\includegraphics[width=8.3cm,angle=0]{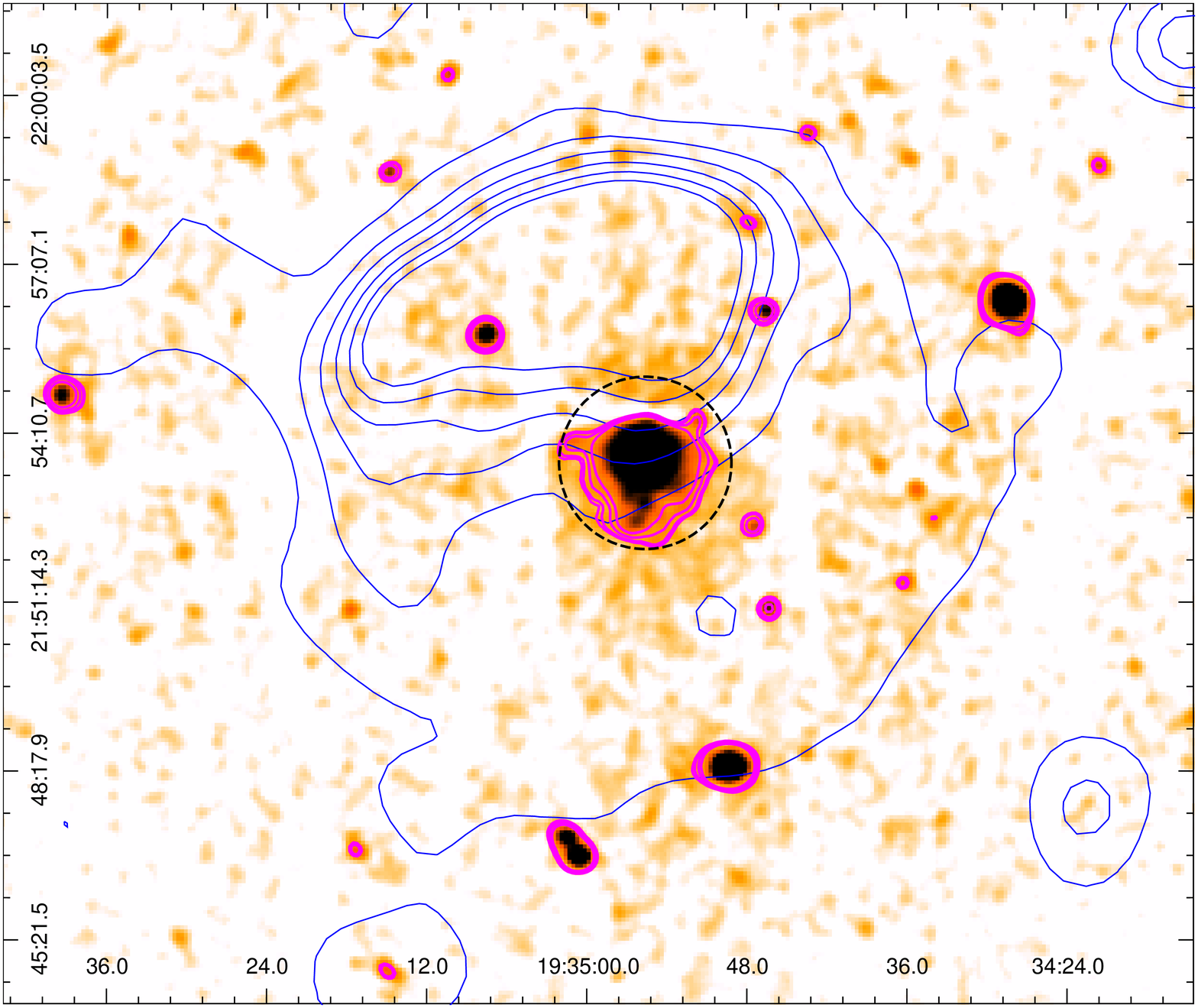}
\includegraphics[width=8.5cm,angle=-90]{profileKallnew.ps}
\caption{\label{ima_psf} Top: 98ks-long \xmm\ PN image of the region around \src; the 1.4 GHz radio map of SNR G57.2+0.8  is also shown (blue contours from the VLA Galactic Plane Survey,  \citealt{stil06}; upper image). The \xmm\ image has been smoothed with a Gaussian function with a radius of 4$^{\prime\prime}$ and magenta contours are displayed in order to emphasise the extended emission around \src. The black dashed circle marks a distance of 90$^{\prime\prime}$ from the \src\ position. Bottom:  {\rc 2014 and 2015} \xmm\ and \cxo\ surface brightness (black crosses, {\rc purple squares}  and  blue crosses, respectively) as a function of the distance from  \src\ compared with their Point Spread Functions  (PSF; red lines; lower plot). The ratios between the data and the PSF are plotted in the lowest panel.}
\end{figure}

Upon visual inspection of the X-ray images, it is apparent that \src\
is embedded in a patch of diffuse emission. To assess this in 
detail, we built for each pn observation a radial profile in the
0.4--10\,keV band and fit a point spread function (approximated by a
King model; \citealt{read11}) to it. In each instance, the inner part
of the profile can be fit by a King model with usual core radius and
slope values, whereas at radii $\approx$30--40$''$ the data start to
exceed significantly the model prediction. Since we obtained
consistent results from all the 2014 observations, we repeated the same
analysis on the stacked images in order to improve the signal-to-noise ratio
of the data. We also selected the photons in the 1--6\,keV energy
range, since the spectral analysis (see Section\,\ref{specana}) shows
that the diffuse emission is more prominent in this band. The combined
2014 \xmm\ profile is shown in black in Fig.\,\ref{ima_psf}. The diffuse emission  emerges
at $\ga$30$''$ from \src\ and extends to at least 70$''$. It is
however not possible to determine where the feature ends, because of
both the low-signal to noise at large distance from the point source
and the gaps between the CCDs. {\rc The profile of the latest \xmm\ dataset 
has been obtained separately from the remaining datasets in order to look 
for shape variabilities of the diffuse component on long timescales. The two pn 
profiles are in agreement within the uncertainties (determined by using a 
Kolmogorov--Smirnov test that there is a substantial probability (>50\%) that 
the two profiles have been extracted from the same distribution), though 
a possible shift of the diffuse component, towards larger radii, might be 
present in the 30$^{\prime\prime}$-\,40$^{\prime\prime}$ radius interval.}

A similar analysis was carried out by using the longest \cxo\ dataset. Though the latter is in CC mode, the field is not particularly crowded and only faint point-like objects are detected in the field of view. Correspondingly, it is still possible to gather information over smaller scales  than in the \xmm\ data. the ACIS-S PSF was simulated using the Chandra Ray Tracer (ChaRT) and Model of AXAF Response to X-rays (MARX v5.0.0-0) software packages\footnote{For more details on the tasks see http://cxc.harvard.edu/chart/index.html and http://space.mit.edu/cxc/marx/index.html}. The result of this analysis is  shown in blue in Figure\,\ref{ima_psf}. Diffuse emission is clearly present in the \cxo\ data and starts becoming detectable at a distance of $>$1$^{\prime\prime}$ from the source. Due to poor statistics we have no meaningful information at radii larger than $\sim$15$^{\prime\prime}$. Therefore, we are not able to assess if the diffuse structures  detected by \xmm\ and \cxo\ are unrelated to each other or linked somehow. 

\begin{table}
\centering \caption{Timing results.} \label{timing}
\begin{tabular}{ll}
\hline
Epoch $T_0$ (MJD)    	                & 56926.0          \\
Validity range (MJD) 	                & 56853.6 -- 56976.4     		 \\
$P(T_0)$ (s)         		        & 3.2450650(1)         	\\
$\dot{P}(T_0)$       	 	        & $1.43(1)\times10^{-11}$ 	     \\
$\ddot{P}(T_0)$ (s$^{-1}$)	                & $-3.5(7)\times10^{-19}$    	 \\
$\nu(T_0)$ (Hz)      		        & 0.30816023(1)       \\
$\dot{\nu}(T_0)$ (Hz s$^{-1}$)           &  $-1.360(3)\times10^{-12}$              \\
$\ddot{\nu}(T_0)$ (Hz s$^{-2}$)               & $3.3(7)\times10^{-20}$ 	  \\
rms residual	(ms)		        & 55 				 \\		
$\chi_\nu^2$ (d.o.f.)		        & 0.57 (6)    			 \\
\hline
$B_p$ (Gauss) &  $2.2\times10^{14}$    \\
$\tau_c$ (yr)  & $3600$    \\
$L_{sd}$ (\lum)  & $1.7\times10^{34}$ \\
\hline
\end{tabular}
\end{table}

\subsection{Timing analysis}\label{timingan}
\begin{figure*}
\centering
\includegraphics[width=8cm,angle=-90]{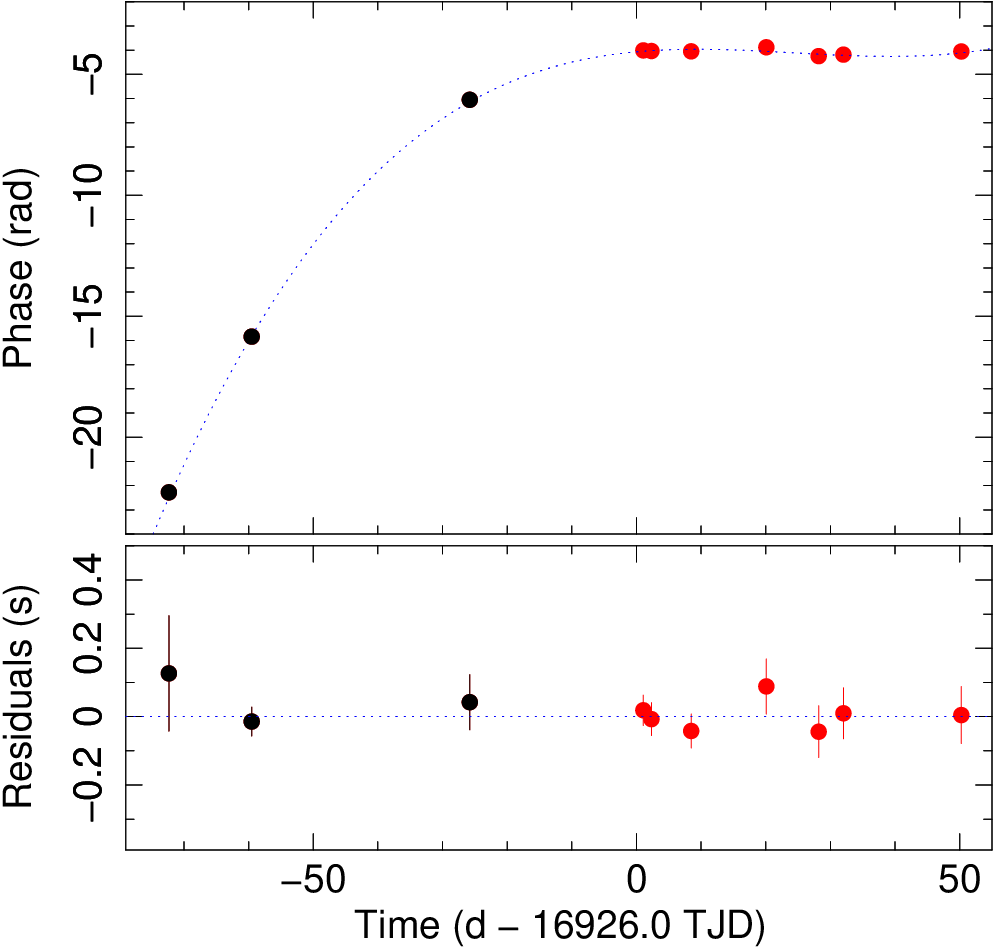}
\hspace{7mm} 
\includegraphics[width=8cm,angle=-90]{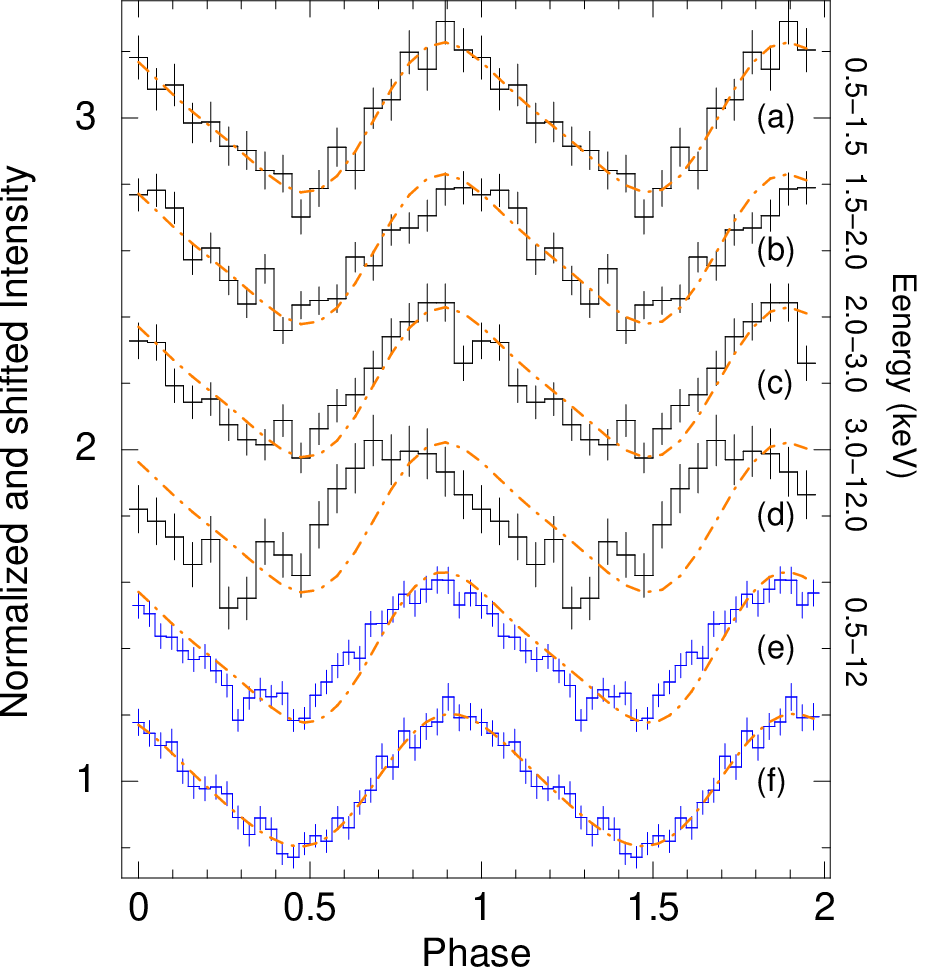}
\caption{\label{timingefold} Left: \src's phase evolution as a function of time fitted with a a linear plus a quadratic plus a cubic components (upper panel). The residuals with respect to our best phase-coherent solution are reported in the lower panel, in units of seconds. Black and red dpoints mark the \cxo\ and \xmm\ observations, respectively. Right: \cxo\ plus \xmm\ background-subtracted pulse profiles (arbitrary shifted on the y-axis). From top to bottom they refer to: (a) 0.5-1.5 keV, (b) 1.5-2.0 keV, (c) 2.0-3.0 keV, (d) 3.0-12.0 keV and (e) 0.5-12.0 keV. The dashed orange  curve marks the best fit (by assuming a model with two sinusoids) of profile (a): a systematic shift towards smaller phases (advance in time) as a function of energy is evident. Profile (f) has been obtained by aligning profiles from (a) to (d).}
\end{figure*}

The 0.5-10\,keV events were used to study the timing properties of the pulsar. The average count rate obtained from \cxo\ and \xmm\ was 0.11$\pm$0.02 cts s$^{-1}$ and 0.21$\pm$0.01 cts s$^{-1}$, respectively.  Coherent pulsations at a period of about 3.24\,s were first discovered in the 2014 July 29 \cxo\ dataset carried out in CC mode \citep{israel14}. The pulse shape is nearly sinusoidal and does not show variations as a function of time. Also the pulsed fraction, defined as the semi-amplitude of the sinusoid divided by the source average count rate, is time independent (within uncertainties) and in the $17\% \div 21\%$ range (1\,$\sigma$ uncertainty of about 1.5\%). Additionally, the pulse shape does not depend on the energy range, though a shift in phase of about 0.16 cycles is clearly detected between the soft (0.5-1.5\,keV) and hard (3.0-12.0\,keV) energy bands, with hard photons anticipating the soft ones (see Figure\,\ref{timingefold}). 

A refined value of $P$=$3.244978(6)$ s (1\,$\sigma$ confidence level; epoch 56866.0 MJD) was inferred based on a phase-coherent analysis. Due to the long time elapsed between the epoch of the first period determination and those of the other \cxo\ observations  we were not able to further extend the timing solution based on the \cxo\ data. Therefore, we inferred a new phase-coherent solution by means of the seven \xmm\ pointings carried out between the end of September and mid November 2014 (red filled circles in left panel of Figure \ref{timingefold}). The new solution also included a first period derivative component: $P$=$3.2450656(2)$ s and $\dot{P}$=$1.37(3)\times$10$^{-11}$ s s$^{-1}$ (1\,$\sigma$ confidence level; epoch 56926.0 MJD; $\chi^2$ of 3.1 for 4 degree of freedom). 

The latter timing solution was accurate enough to include the previous \cxo\ pointings (black filled circles in left panel of Figure \ref{timingefold}). The final timing solution, encompassing the whole dataset, is reported in Table \ref{timing} and includes a second period derivative acting in the direction of decelerating the rate of period change $\dot{P}$. The inclusion of the new $\ddot{P}$ component has a F-test probability of $8\times10^{-4}$ and $10^{-7}$ of not being needed (when considering only the \xmm\ datasets or the whole ten pointings in the fit, respectively). Moreover, the new timing solution implies a r.m.s. variability of only 55 ms, corresponding to a timing noise level of less than 2\%, well within the value range observed in isolated neutron stars. 

We note that the second period derivative we found is  unlikely to result from a change, as a function of time, of the pulse profiles, which are almost sinusoidal and show no evidence for variation (see right panel of Figure \ref{timingefold}). 

We notice that this analysis is valid under the assumption that the location and geometry of the emitting region remains constant throughout the observations, as suggested by studies of other transient magnetars (see \citealt{perna08,albano10}). 

{\rc The accuracy of the timing solution reported in Table\,\ref{timing} is not good enough to coherently include the March 2015 \xmm\ data. Correspondingly, we inferred the period for this latest pointing similarly to what reported above finding a best value of  $P$=$3.24528(6)$ s (95\% confidence level; epoch 57106.0 MJD). This is less than 2$\sigma$ away from the 
expected period extrapolated from the timing solution in Table\,\ref{timing}. The pulse profile parameters changed significantly with respect to the previous datasets with a pulsed fraction of only 5$\pm$1\% (1$\sigma$) and a more asymmetric  shape.}

\subsection{Spectral analysis}\label{specana}
For the phase-averaged spectral analysis (performed with \textsc{xspec 12.8.2} fitting package; \citealt{arnaud96}) we started by considering all the datasets together. Then, we concentrated on the 29 July 2014 data, being the longest and highest statistics \cxo\ pointing (about 75ks effective exposure for 8200 photons) and the \xmm\ pn spectra (effective exposure time of about {\rc 105ks and 22000 events}). 
A summary of the spectral fits is given in Table\,\ref{specs}. To account for the above reported diffuse component (see Section\,\ref{diffuse}) we used, as background spectra of the point-like central source, the regions we described in Section\,\ref{acisdata} and \ref{xmmdata} and from which we extracted later the diffuse component spectra. 

We started by fitting all the 10 datasets {\rc carried out during 2014} separately  leaving free to vary all the parameters. The absorption was forced to be free but the same among observations. Photons having energies below about 0.8\,keV and above 10\,keV were ignored, owing to the very few counts from \src\ (energy channels were rebinned in a way of having at least 30 events). Furthermore, all the energy channels consistent with zero after the background subtraction were ignored.  The abundances used were those of \citet{wilms00}.
The spectra were not fitted well by any single component model such as a power-law (PL) or blackbody (BB) which gave a reduced $\chi^2$ in the 1.2 -- 1.8 range depending on the used single component (282 and 407 degrees of freedom, hereafter d.o.f., for the \cxo\ and \xmm\ spectra, respectively).  A canonical two-component model often used  to model magnetars, i.e. an absorbed BB plus PL, resulted in a good fit with reduced $\chi^2$ of 0.99 (280 d.o.f.) and 1.03 (405 d.o.f.) for the \cxo\ and \xmm\ spectra, respectively. The inclusion of a further spectral component (the BB in the above procedure) was evaluated to have a formal F-test probability equal to 4.5\,$\sigma$ and 7.0\,$\sigma$ (for \cxo\ and \xmm, respectively) of being significant.  

\begin{figure}
\centering
\resizebox{\hsize}{!}{\includegraphics[angle=-90]{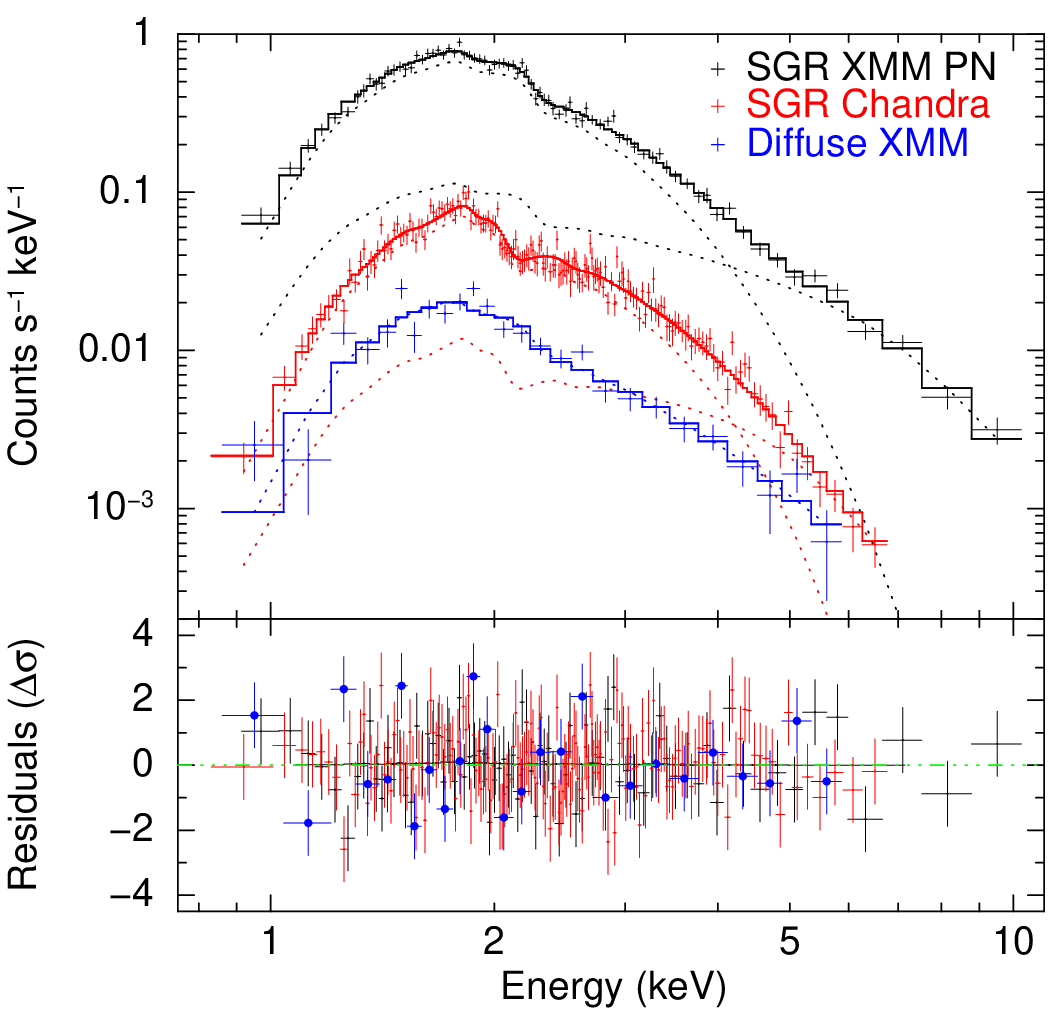}}
\resizebox{\hsize}{!}{\includegraphics[angle=-90]{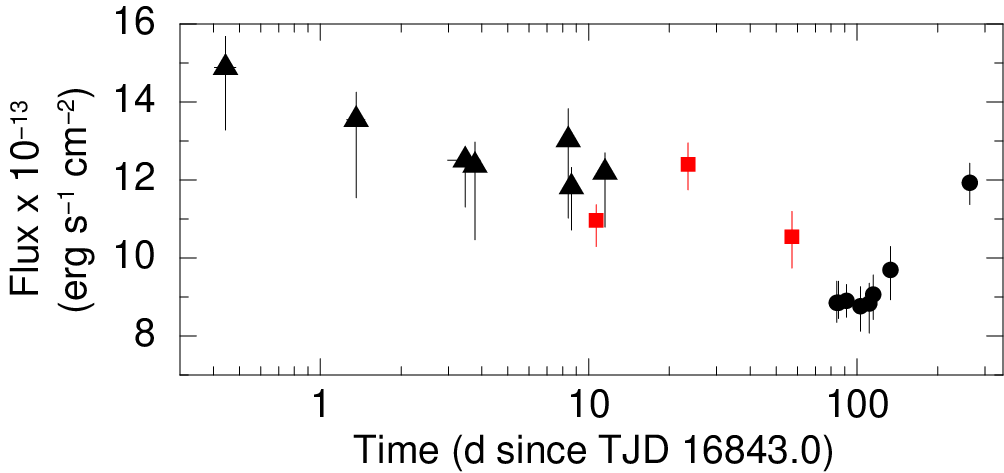}}
\caption{\label{specall} Spectra of \src\ and of the diffuse  emission around the pulsar. From  top to  bottom: \src\ cumulative \xmm\ PN spectrum, the \src\ \cxo\ ACIS spectrum of observation 15875 and the  cumulative \xmm\ PN spectrum of the diffuse emission (upper plot). Residuals (is $\sigma$ units) are shown and refer to the absorbed PL+BB model for \src\ and to a PL model for the diffuse component. {\rc Time evolution for the absorbed 1-10 keV flux of \src\ obtained by using datasets from \swift\ (black triangles), \cxo\ (red squares) and \xmm\ (black circles). The zero on the x-axis marks the \swift\ BAT trigger.}}
\end{figure}

A  flux variation, of the order of about 25\%, was clearly detected between the \cxo\ and \xmm\ {\rc 2014} pointings. On the other hand no significant  flux variation was detected among spectra of  \xmm\ observations. Correspondingly, in order to increase the statistics we proceeded to combine the seven \xmm\ {\rc 2014} spectra together (we used   the \textsc{sas} task \textsc{epicspeccombine}). By using the latter spectrum we obtain  a F-test probability of 7.8\,$\sigma$ that the BB component inclusion is significant.
In {\rc the upper panel of} Figure\,\ref{specall} the \xmm\ combined source spectrum (in black) is reported together with the \cxo\ spectrum of the longest pointing (in red; the two further \cxo\ spectra are not shown in Figure for clarity purposes). We note that, within about 1\,$\sigma$ uncertainties, the \cxo\ and \xmm\ spectral parameter are consistent with each other with the exception of the flux. 

{\rc The latest \xmm\ pointing, carried out in March 2015, was not combined with the previous ones in order to look for spectral variability on long time scales. While the PL plus BB  spectral decomposition holds also for this dataset, the flux significantly increased by about 25\% reaching a level similar to that of the longest \cxo\ pointing in July 2014. It is evident from Table\,\ref{specs} that the only significantly changed parameter is the flux of the PL component.

Due to the poor statistics of the \swift\ XRT spectra we only inferred the 1-10keV fluxes by assuming the PL plus BB model obtained by the combined \xmm\ spectrum and including  a scale factor which was free to vary in order to track the flux variation through the outburst. 
The lower panel of Figure\,\ref{specall} includes all the 1-10keV observed fluxes inferred from the \swift, \cxo\ and \xmm\ spectra. It is evident that the source is still variable above a general decay trend.}

\begin{table*}
\begin{minipage}{17cm}
\centering \caption{\cxo\ and \xmm\ spectral results. Errors are at a 1\,$\sigma$ confidence level for a single parameter of interest.} \label{specs}
\begin{tabular}{@{}lccccccc}
\hline
Mission (Model) & \nh$^b$ & $\Gamma$ & $kT$   & $R_{\mathrm{BB}}$ $^c$ & Flux$^d$ & Luminosity$^d$ & $\chi^2_\nu$ (dof) \\
 & ($10^{22}$ cm$^{-2}$) &  & (keV)  & (km) & {($10^{-12}$ \flux)}& ($10^{34}$ \lum) & \\
\hline
\multicolumn{8}{c}{\em SOURCE EMISSION} \\
\hline
\textsc{CHANDRA (BB + PL)} & $2.0\pm0.4$ & $2.8\pm0.8$ & $0.45\pm0.03$  & $1.9\pm0.2$  & $1.24\pm0.06$ & $3.1\pm0.5$ & 0.97 (165) \\

\textsc{XMM (BB + PL)} & $1.6\pm0.2$ & $1.8\pm0.5$ & $0.47\pm0.02$  &  $1.6\pm0.1$ & $0.89\pm0.05$ & $1.7\pm0.4$ & 1.02 (74) \\
\textsc{\rc XMM$^e$ ~~~~~~"} & ${\rc 1.6\pm0.2}$ &  ${\rc 2.1\pm0.4}$ & ${\rc 0.48\pm0.02 }$ & ${\rc 1.6\pm0.2 }$ & ${\rc 1.19\pm0.06 }$& ${\rc 2.4\pm0.5}$ & {\rc 0.93 (109)} \\
\hline
\multicolumn{8}{c}{\em DIFFUSE EMISSION} \\
\hline
\textsc{XMM (PL)} & $3.8\pm0.4$ & $3.8\pm0.3$ & $--$  & $--$  & $0.14\pm0.02$ & $0.6\pm0.1$ & 1.94 (23) \\
\hline
\end{tabular}
\begin{list}{}{}
\item[$^{a}$] \textsc{xspec} models; \textsc{bb = bbodyrad, pl = powerlaw}. 
\item[$^{b}$]  We used the abundances of \citet*{wilms00} 
\item[$^{c}$] The blackbody radius is calculated at infinity and for an arbitrary distance of 9\,kpc.
\item[$^{d}$] In the 1--10\,keV energy band; fluxes are observed values, luminosities 
are de-absorbed quantities.
\item[$^{e}$] {\rc March 2015 \xmm\ observation.}
\end{list}
\end{minipage}
\end{table*}

The same background regions used to correct the EPIC pn source spectra were then assumed as a reliable estimate of the diffuse emission. For the background of the diffuse emission we considered two regions laying far away (at a distance $>$4$^{\prime}$) from the pulsar and in two different CCDs obtaining similar results in both cases. We first fit all the seven spectra together. The use of one spectral component gave a relatively good fit with a reduced $\chi^2$ of 1.22 and 1.33 (107 d.o.f.) for an absorbed PL and BB model, respectively. Then we left free to vary all the parameters resulting in a  reduced $\chi^2$ of 1.15 and 1.18 (95 d.o.f.) for the PL and BB model, respectively. While no improvement was achieved for the BB model the PL model appears to vary among \xmm\ observations at about 2.0\,$\sigma$ confidence level. Therefore, we conclude that there is no suggestion of variability for the diffuse emission. A combined (from the seven \xmm\ pointings) spectrum for the diffuse emission was obtained, in a way similar to that already described for the source spectrum. The \xmm\ combined spectrum of the diffuse emission and the results of the spectral fitting for the PL model are shown in Figure\,\ref{specall} and in Table\,\ref{specs}. Two facts can be immediately evinced: a simple model is not a good approximation for the diffuse emission and the absorbing column is significantly different from the one we inferred for the magnetar. At present stage we cannot exclude that the two things are related to each other. In particular, we note that the largest values of the residuals originated from few ``random" datapoints rather than by an up-and-down trend (often suggesting a wrong adopted continuum component; see blue points in the lower panel of Figure\,\ref{specall}). {\rc Also for the diffuse emission we kept separated the 2015 \xmm\ observation in order to look for spectral variations. Unfortunately, the low statistics prevented us in checking if changes in the spectral parameters are present. The inferred 1-10keV observed flux is  (1.67$\pm_{\rm \bf 0.05}^{\rm \bf 0.03}$)$\times$10$^ {\rm -13}\flux$, in agreement with the 2014 value. }

\subsection{Pre-outburst observations} 

\swift\ XRT observed \src\ twice before its activation during the
\swift\ Galactic plane survey (see \citealt{campana14}).
The first observation took place on Dec 30, 2010 for 514\,s (obsid
00045278001). \src\ is far off-axis ($\sim$10$^{\prime}$) and 
we derived  a $3\,\sigma$ upper limit
of $3.2\times 10^{-2}$ cts s$^{-1}$ .

The second observation took place on Aug 28, 2011 for 617\,s (obsid
00045271001). \src\ is detected at a rate $(1.55\pm0.63)
\times 10^{-2}$ cts s$^{-1}$.
Assuming the same spectral model of the \xmm\ observations (see Section \ref{specana} and Table\,\ref{specall}, we derive
a 1-10\,keV luminosity of $(9.3\pm3.6)\times 10^{33}$ erg s$^{-1}$  (including uncertainties in the count rate and {\rc assuming a distance} of 9\,kpc). 

The field was also imaged during the ROSAT all-sky survey twice, but the high column density prevents any firm upper limit on the observed flux.

\section{Radio observations}
The first radio follow-up observations of \src\ were carried out on 9 and 14 July 2014 from the Ooty Radio Telescope (ORT) and the Giant Meterwave Radio Telescope (GMRT), at 326.5 and 610.0 MHz, respectively \citep{pippone}. No pulsed radio emission was found down to a flux of 0.4 mJy and 0.2 mJy at 326.5 and 610.0 MHz (assuming a 10\% duty cycle), respectively.

The source was observed with the Parkes radio telescope at 10-cm and 20-cm in four epochs between 1 and 3 August, shortly after the detection of X-ray pulsations \citep{israel14}, and again at 10-cm on 28 September, almost simultaneously with one of our \xmm\ observations. Observations at 10-cm were obtained using the \textsc{atnf} Digital Filterbanks \textsc{dfb3} (used in search mode with a sampling time of 1 ms) and DFB4 (in folding mode) at a central frequency of 3100 MHz, over 1024 MHz of bandwidth. 20-cm observations were acquired using the reconfigurable digital backend \textsc{hipsr} (HI-Pulsar signal processor) with a central frequency of 1357 MHz, a 350 MHz bandwidth and a sampling time of 64 $\mu$s. Further details of the observations are summarized in Table \ref{tab:radio}.

Data were folded in 120-s long sub-integrations using the ephemeris in Table\,\ref{timing} and then searched over a range of periods, spanning $\pm 1.5$\,ms with respect to the X-ray value of any given observing epoch, and over dispersion measures (DM) up to 1000 pc$\,$cm$^{-3}$.

The data acquired in search mode were also blindly searched over DMs up to 1000 both for periodic signals and single dedispersed pulses. The 20-cm data were searched in real time using \textsc{heimdall}\footnote{ see http://sourceforge.net/projects/heimdall-astro/ for further details}, while the 10-cm data were analysed with the package \textsc{sigproc} (http://sigproc.sourceforge.net/). No pulsed signal with a period similar to that detected in X-rays, nor single dispersed pulses were found down to a signal-to-noise ratio of 8. Table \ref{tab:radio} lists the upper limits obtained at each epoch and frequency.

\begin{table}
\centering \caption{The table lists for each radio observation: the date and time (UT) of the start of the acquisition (in the form yy-mm-dd-hh:mm); the receiver used, either the 10-cm feed of the coaxial 10-50cm \citep{granet05} or the central beam of the 20-cm multibeam receiver \citep{staveley96}; the integration time; the flux density upper limit for a pulsed signal with a 3.2\,s period; the flux density upper limit for a single pulse of 32\,ms duration. Flux are expressed in mJy units.} \label{tab:radio}
\begin{tabular}{lllll}
\hline
UT Start & Rec & T$_{obs}$ (h) & S$_{min}$  & S$_{min}^{sp}$ \\
\hline
14-08-01-11:34 & 10-50cm & 3.0 & 0.04 & 68 \\
14-08-02-11:22 & 10-50cm & 3.0 & 0.04 & 68 \\
14-08-03-12:29 & 20cm-MB & 1.5 & 0.05 & 61 \\
14-08-03-13:32 & 10-50cm & 1.0 & 0.07 & 68 \\
14-09-28-08:34 & 10-50cm & 2.0 & 0.05 & 68 \\
\hline
\end{tabular}
 \end{table}

\section{Discussion}\label{discussion}

Thanks to an intensive \cxo\ and \xmm\ observational campaign of \src\ covering the first {\rc 8} months since the first bursts detected by \swift BAT, we were able to infer the main timing and spectral properties of this newly identified member of the magnetar class. In particular, we discovered  strong coherent pulsations at a period of about 3.24\,s in a \cxo\ long pointing carried out in July 2014. Subsequently, by using the \xmm\ observations (spaced so to keep the pulse phase coherence among pointings) we started building a timing solution by means of a phase fitting technique.  We were able to phase-connect all the {\rc 2014} \cxo\ and \xmm\ datasets and we inferred both a first and second period derivative. 
These findings further confirm that \src\ is indeed a magnetar which is slowing-down at a rate of about half a millisecond per year. However, this trend is slowing-down due to a negative   $\ddot{P}$ (see Table\,\ref{timing}). The accurate timing solution allowed us also to infer the dipolar magnetic field strength, an upper limit on the {\rc true pulsar age} and the corresponding spin-down luminosity (under usual assumptions).

\src\ is a seemingly young object, $\leq 3$\,kyr, with a $B_p$ value ($\sim$2.2$\times$10$^{14}$\,Gauss) well within the typical range of magnetars. The X-ray emission is pulsed. The pulse shape is energy independent (within uncertainties) and it is almost sinusoidal with 
a $\sim 20\%$ pulsed fraction (measured as the semi-amplitude of the sinusoid divided by the average count rate) {\rc during 2014. It becomes less sinusoidal with a pulsed fraction of only 5\% during the latest \xmm\ observation}. We detected an 
energy-dependent phase shift ($\sim$0.16 cycles at maxiumum), with the hard photons anticipating the soft ones. 
This behaviour is not very common among known magnetars, \rxs\  being a notable exception (though with a different trend in energy; see \citealt{israel01,roz05}). In \rxs\ the shift is likely associated to the presence of a (spin phase) variable hard X-ray component extending up to at least 100\,keV \citep{kuiper06, gri07}.  Similarly, the pulse profile phase shift of \src\ might be due to the presence of at least two distinct components (peaks) with different weight at different energies. 
The non  detection of emission from \src\ at energies above 10\,keV does not allow us to firmly assess the cause of the shift. 

The source spectrum can be well described by the canonical two-component model often applied to magnetars, i.e. an absorbed black body plus a power-law (kT$\sim$0.5\,keV and $\Gamma\sim$2). The \src\ 1-10\,keV observed flux of 1.5$\times$10$^{-12}$ erg cm$^{-2}$ s$^{-1}$ is among the  lowest observed so far from magnetars at the beginning of their outbursts. 
Although it is possible that we missed the outburst onset (which perhaps occurred before the first burst epoch), a backward search of burst activity in the BAT data at the position of \src\ gave negative results \citep{cummings14b}. Emission from \src\ is detected in an archival \swift\ XRT pointing in 2011 at a flux only a factor of few lower than the one detected soon after the burst emission.  At current stage we cannot exclude that the source has not reached the quiescent level or that it has a relatively bright quiescent luminosity. {\rc This latter possibility is partially supported by the unusual properties of \src\ which displays both intervals of flux weakening a brightening superimposed to a slow decay. We note that the latest \xmm\ pointing occurred less than 20 days from the Konus-{\it Wind} detection of the first intermediate flare from this source \citep{golenetskii15}. } 

A significant diffuse emission, extending from spatial scales of $>$1$^{\prime\prime}$ up to more than 1$^{\prime}$ around the magnetar, was clearly detected both by \cxo\ and \xmm. Due to the use of different instruments/modes at different epochs we were not able to test if the diffuse component varied in time (as expected in the case of scattering by dust clouds on the line-of-sight) between the \cxo\ and \xmm\ pointings. Among the \xmm\ pointings the component does not change significantly. The \cxo\ data allowed us to sample the spatial distribution of the component only up to about 20$^{\prime\prime}$ (at larger radii we are hampered by the statistics), while the lower spatial resolution of the \xmm\ pn allowed us to detect the diffuse emission only beyond about 20$^{\prime\prime}$. 
{\rc We do not detect any flux variation for the diffuse emission among the eight \xmm\ pointings despite the pulsar enhancement of about 20\% between October 2014 and March 2015, a result which would favour a magnetar wind nebula (MWN) interpretation.}   
The PL model used to fit the pn spectra implies a relatively steep photon index of about 3.8 which is similar to what observed for the  candidate MWN around \sgrj\ \citep{younes12}, but at the same time is steeper than the PL photon index of \src\ suggesting that the dust scattering scenario might be more likely.

In \sgrj\ two diffuse components have been identified: a symmetric component around the magnetar extending up to about 50$^{\prime\prime}$ interpreted as a dust scattering halo \citep{younes12,etr13}, and an asymmetric component extending up to 150$^{\prime\prime}$ proposed as a wind nebula \citep{younes12}. The spectrum of the former component has a PL photon index steeper than that of the magnetar (which however, at variance with \src, is fitted well by a single PL alone likely due to a very high absorption which hampers the detection of any soft BB), while the latter has a flatter spectrum. In order to compare the properties of the diffuse emission around \sgrj\ and \src, we  fitted the  
\cxo\ and  \xmm\  spectra of \src\ with a PL alone obtaining a photon index of 4.4$\pm$0.1 and 4.3$\pm$0.1 (we used only photons in the 1.5-8.0\,keV band similar to the case of \sgrj)  implying that the diffuse component might have a spectrum flatter than that of the magnetar  and favouring the wind nebula scenario. In the latter case {\rc the
efficiency at which the rotational energy loss of a pulsar, \.E$_{\rm \rc rot}$,
is radiated by the PWN is given by ${\rm \rc \eta_X}$ =  ${\rc L}_{\rm \rc X,pwn}$/\.E$_{\rm rot}$ = }(0.6$\times$10$^{34}$/1.7$\times$10$^{34}$)$\simeq$0.35, not that different from what inferred from similar components around \sgrj\ and \rrata\ \citep{younes12,rea09b}.  Further \xmm\ and/or \cxo\ observations taken at flux levels significantly   different from those we recorded so far should help in settling the nature of the diffuse emission.

A search for radio pulsed emission from \src\ gave negative result down to a  flux density of about 0.5 mJy (and 70 mJy for a single pulse). It has been suggested that whether or not a magnetar can also shine as a transient radio pulsar might depend on the ratio between its quiescent X-ray luminosity and spin-down luminosity, given that all magnetars with detected radio pulsed emission have this ratio smaller than $\sim$0.3 \citep{rea12}, at variance with typical radio-quiet magnetars that have quiescent X-ray luminosity normally exceeding their rotational power. Based on the coherent timing solution we inferred a spin-down luminosity of about 2$\times$10$^{34}$ \lum. At the present stage it is also rather difficult to obtain a reliable value of the  quiescent luminosity due to the uncertainties on the distance and the flux of the \swift\ pre-burst detection. If a distance of 9\,kpc is assumed, the \swift\ faintest flux convert to a luminosity of about  5$\times$10$^{33}$ \lum which results in L$_X$/L$_{sd}$ $\sim$ 0.25, close to 0.3 limiting value. However, if the distance is larger and/or  the quiescent flux is a factor of few larger than estimated from \swift, the source would move toward higher values of L$_{X,qui}$/L$_{sd}$ in the ``radio-quiet" region of the fundamental plane (see left panel of Figure\,2 in \citealt{rea12}). Correspondingly, the non detection of radio pulsations might be not that surprising.

The uncertainty in the quiescent level of this new magnetar makes any
attempt to infer its evolutionary history rather uncertain. Given the short characteristic age (a few kyrs, which is most probably 
representative of the true age given that no substantial field decay is 
expected over such a timespan), the present value of the magnetic field is likely  
not that  different from that at the moment of birth.
The above reviewed timing characteristics would then be consistent with a
quiescent bolometric luminosity of the order of
$\sim5\times10^{33-34}$\,erg~s$^{-1}$ (see Fig.\,11 and 12 in \citealt{vigano13}), depending on the assumed magnetic field geometry and envelope composition.

Constraints on its outburst luminosity evolution can be put from general considerations (see \citealt{pons12,vigano13}). 
If we assume that the flux derived by the pre-outburst \swift\ observations 
provides a correct estimate of the magnetar quiescence, and we rely on a distance of 9\,kpc, then the source luminosity increases
from a quiescent level of $L_{X, qui}\sim7\times10^{33}$\,erg~s$^{-1}$
to a 'detected' outburst peak  of $L_{X,
out}\sim4\times10^{34}$\,erg~s$^{-1}$. Such luminosity variation within the outburst (about a
factor  of 5) is rather small for a magnetar with a medium-low
quiescent level (see Fig.\,2 of \citealt{pons12}). 
In particular, the outburst peak luminosity usually reaches
about $L_{X, out}\sim5\times10^{35}$\,erg~s$^{-1}$, due to the typical
energies released in magnetars' crustal fractures (about
$10^{44-45}$\,erg; \citealt{pons12,perna11}), coupled with estimates of the
neutrino cooling efficiencies \citep{pons12}. If there are no
intrinsic physical differences between this outburst and other
magnetar outbursts (see \citealt{rea11}), then we
can foresee two possibilities to explain the relatively low maximum luminosity detected.

The first possibility is that we have missed the real outburst peak of \src, 
which was then caught already during its outburst decay.
In this case the quiescent luminosity claimed by the archival \swift\ 
observation might be correct, and the magnetar had a flux
increase during the outburst, but we could catch it only thanks to
an SGR-like burst detected when the magnetar had already cooled down
substantially. Given the typical outburst cooling curves, we can
roughly estimate that in this scenario we observed the source about 10-40\,days after its real
outburst onset.

The second possibility is that the source distance is farther than the assumed SNR
distance of 9\,kpc (note that the method used by \citealt{pavlovic13} to infer this distance implies a relatively large degree of uncertainty, even a factor of two in both directions). To have an outburst peak luminosity in line with other magnetars, \src\ should have a
distance of $\sim$20-30\,kpc. At this distance the assumed \swift\, quiescence level
would also be larger ($\sim7\times10^{34}$\,erg~s$^{-1}$), hence a
factor of $\sim$5 in increase in luminosity in the outburst would then
be in line with what observed (and predicted) in other cases (see
again Fig.\,2 of \citealt{pons12}). However, in the direction of
\src\, the Galaxy extends until $\sim$14\,kpc \citep{hou09} making
such a large distance rather unlikely.

We then suggest that the very low peak flux of the detected outburst
of \src\, has no different physics involved with respect
to other magnetar outbursts, but we have simply missed the onset of
the outburst. If the flux detected by \swift\  before the outburst was its quiescent level, we envisage that the outburst onset occurred about a month before the first X-ray burst detection. If future observations will set the source at a lower quiescent level,  the outburst peak should have occurred even longer before we first detected its activity.


\section*{Acknowledgements} 
The scientific results reported in this article are based on observations obtained with the {\em Chandra X-ray Observatory} and {\xmm}, an ESA science mission with instruments and contributions directly funded by ESA Member States and NASA. This research has made use of software provided by the {\cxo} X-ray Center (CXC) in the application package CIAO, and of softwares and tools provided by the High Energy Astrophysics Science Archive Research Center (HEASARC), which is a service of the Astrophysics Science Division at NASA/GSFC and the High Energy Astrophysics Division of the Smithsonian Astrophysical Observatory. {\rc This research is based on observations with the NASA/UK/ASI \swift\ mission.}
We thank N. Schartel for approving the \xmm\ November 2014 observation through the Director's Discretionary Time program and the staff of the \xmm\ Science Operation Center for performing the Target of Opportunity observations. Similarly, we thank B. Wilkes for approving the \cxo\ August 2014 observation through the Director's Discretionary Time program and the \cxo\ staff for performing the Target of Opportunity observations. 
{\rc We thank the \swift\ duty scientists and science planners for making these
observations possible.} The Parkes radio telescope is part of the Australia Telescope which is funded by the Commonwealth of Australia for operation as a National Facility managed by CSIRO. The authors warmly thank Phil Edwards for the prompt scheduling of the observations and John Reynolds for releasing part of the telescope time of his project. We thanks Jules Halpern for useful comments. NR is supported by an NWO Vidi
Grant, and by grants AYA2012-39303 and SGR2014-1073. This work is partially supported by the European COST Action MP1304 (NewCOMPSTAR).

\bibliographystyle{mn2e}
\bibliography{biblio}

\begin{thebibliography}{}

\bibitem[\protect\citeauthoryear{{Albano}, {Turolla}, {Israel}, {Zane},
  {Nobili} \& {Stella}}{{Albano} et~al.}{2010}]{albano10}
{Albano} A.,  {Turolla} R.,  {Israel} G.~L.,  {Zane} S.,  {Nobili} L.,
  {Stella} L.,  2010, \apj, 722, 788

\bibitem[\protect\citeauthoryear{{Arnaud}}{{Arnaud}}{1996}]{arnaud96}
{Arnaud} K.~A.,  1996, in {Jacoby} G.~H.,  {Barnes} J.,  eds, Astronomical Data
  Analysis Software and Systems V Vol.~101 of Astronomical Society of the
  Pacific Conference Series, {XSPEC: The First Ten Years}.
ASP, San Francisco, pp 17--20

\bibitem[\protect\citeauthoryear{{Burgay}, {Israel}, {Rea}, {Possenti},
  {Zelati}, {Esposito}, {Mereghetti} \& {Tiengo}}{{Burgay}
  et~al.}{2014}]{burgay14}
{Burgay} M.,  {Israel} G.~L.,  {Rea} N.,  {Possenti} A.,  {Zelati} F.~C.,
  {Esposito} P.,  {Mereghetti} S.,    {Tiengo} A.,  2014, The Astronomer's
  Telegram, 6371, 1

\bibitem[\protect\citeauthoryear{{Burrows}, {Hill}, {Nousek}, {Kennea},
  {Wells}, {Osborne}, {Abbey}, {Beardmore}, {Mukerjee}, {Short}, {Chincarini},
  {Campana}, {Citterio}, {Moretti}, {Pagani}, {Tagliaferri}, {Giommi},
  {Capalbi}, {Tamburelli} \& {Angelini}}{{Burrows} et~al.}{2005}]{burrows05}
{Burrows} D.~N.,  {Hill} J.~E.,  {Nousek} J.~A.,  {Kennea} J.~A.,  {Wells} A.,
  {Osborne} J.~P.,  {Abbey} A.~F.,  {Beardmore} A.,  {Mukerjee} K.,  {Short}
  A.~D.~T.,  {Chincarini} G.,  {Campana} S.,  {Citterio} O.,  {Moretti} A.,
  {Pagani} C.,  {Tagliaferri} G.,  {Giommi} P.,  {Capalbi} M.,  {Tamburelli}
  F.,    {Angelini} L., 2005 , Space Sci. Rev., 120, 165


\bibitem[\protect\citeauthoryear{{Camero-Arranz}, {Rea}, {Bucciantini},
  {McLaughlin}, {Slane}, {Gaensler}, {Torres}, {Stella}, {de O{\~n}a},
  {Israel}, {Camilo} \& {Possenti}}{{Camero-Arranz} et~al.}{2013}]{camero13}
{Camero-Arranz} A.,  {Rea} N.,  {Bucciantini} N.,  {McLaughlin} M.~A.,  {Slane}
  P.,  {Gaensler} B.~M.,  {Torres} D.~F.,  {Stella} L.,  {de O{\~n}a} E.,
  {Israel} G.~L.,  {Camilo} F.,    {Possenti} A.,  2013, \mnras, 429, 2493

\bibitem[\protect\citeauthoryear{{Camilo}, {Ransom}, {Halpern}, {Reynolds},
  {Helfand}, {Zimmerman} \& {Sarkissian}}{{Camilo} et~al.}{2006}]{camilo06}
{Camilo} F.,  {Ransom} S.~M.,  {Halpern} J.~P.,  {Reynolds} J.,  {Helfand}
  D.~J.,  {Zimmerman} N.,    {Sarkissian} J.,  2006, \nat, 442, 892

\bibitem[\protect\citeauthoryear{{Campana}, {de Ugarte Postigo}, {Thoene},
  {Gorosabel}, {Rea} \& {Coti Zelati}}{{Campana} et~al.}{2014}]{campana14}
{Campana} S.,  {de Ugarte Postigo} A.,  {Thoene} C.~C.,  {Gorosabel} J.,  {Rea}
  N.,    {Coti Zelati} F.,  2014, GRB Coordinates Network, 16535, 1

\bibitem[\protect\citeauthoryear{{Cummings}, {Barthelmy}, {Chester} \&
  {Page}}{{Cummings} et~al.}{2014}]{cummings14}
{Cummings} J.~R.,  {Barthelmy} S.~D.,  {Chester} M.~M.,    {Page} K.~L.,  2014,
  The Astronomer's Telegram, 6294, 1

\bibitem[\protect\citeauthoryear{{Cummings} \& {Campana}}{{Cummings} \&
  {Campana}}{2014}]{cummings14b}
{Cummings} J.~R.,  {Campana} S.,  2014, The Astronomer's Telegram, 6299, 1

\bibitem[\protect\citeauthoryear{{De Luca} \& {Molendi}}{{De Luca} \&
  {Molendi}}{2004}]{deluca04}
{De Luca} A.,  {Molendi} S.,  2004, \aap, 419, 837

\bibitem[\protect\citeauthoryear{{Duncan} \& {Thompson}}{{Duncan} \&
  {Thompson}}{1992}]{duncan92}
{Duncan} R.~C.,  {Thompson} C.,  1992, \apjl, 392, L9

\bibitem[\protect\citeauthoryear{{Esposito}, {Tiengo}, {Rea}, {Turolla},
  {Fenzi}, {Giuliani}, {Israel}, {Zane}, {Mereghetti}, {Possenti}, {Burgay},
  {Stella}, {G{\"o}tz}, {Perna}, {Mignani} \& {Romano}}{{Esposito}
  et~al.}{2013}]{etr13}
{Esposito} P.,  {Tiengo} A.,  {Rea} N.,  {Turolla} R.,  {Fenzi} A.,  {Giuliani}
  A.,  {Israel} G.~L.,  {Zane} S.,  {Mereghetti} S.,  {Possenti} A.,  {Burgay}
  M.,  {Stella} L.,  {G{\"o}tz} D.,  {Perna} R.,  {Mignani} R.~P.,    {Romano}
  P.,  2013, \mnras, 429, 3123

\bibitem[\protect\citeauthoryear{{Golenetskii}, {Aptekar}, {Pal'Shin},
  {Frederiks}, {Svinkin}, {Cline}, {Hurley}, {Mitrofanov}, {Golovin}, {Litvak},
  {Sanin}, {von Kienlin}, {Zhang}, {Rau}, {Savchenko}, {Bozzo}, {Ferrigno},
  {Boynton}}{{Golenetskii} et~al.}{2015}]{golenetskii15}
{Golenetskii} S.,  {Aptekar} R.,  {Pal'Shin} V.,  {Frederiks} D.,  {Svinkin}
  D.,  {Cline} T.,  {Hurley} K.,  {Mitrofanov} I.~G.,  {Golovin} D.,  {Litvak}
  M.~L.,  {Sanin} A.~B.,  {von Kienlin} A.,  {Zhang} X.,  {Rau} A.,
  {Savchenko} V.,  {Bozzo} E.,  {Ferrigno} C.,  {Boynton} W.,  {Fellows} C.,
  {Harshman} K.,  {Enos} H.,    {Starr} R.,  2015, GRB Coordinates Network,
  17699

\bibitem[\protect\citeauthoryear{{G{\"o}tz}, {Rea}, {Israel}, {Zane},
  {Esposito}, {Gotthelf}, {Mereghetti}, {Tiengo} \& {Turolla}}{{G{\"o}tz}
  et~al.}{2007}]{gri07}
{G{\"o}tz} D.,  {Rea} N.,  {Israel} G.~L.,  {Zane} S.,  {Esposito} P.,
  {Gotthelf} E.~V.,  {Mereghetti} S.,  {Tiengo} A.,    {Turolla} R.,  2007,
  \aap, 475, 317

\bibitem[\protect\citeauthoryear{{Granet}, {Zhang}, {Forsyth}, {Graves},
  {Doherty}, {Greene}, {James}, {Sykes}, {Bird}, {Sinclair}, {Moorey} \&
  {Manchester}}{{Granet} et~al.}{2005}]{granet05}
{Granet} C.,  {Zhang} H.~Z.,  {Forsyth} A.~R.,  {Graves} G.~R.,  {Doherty} P.,
  {Greene} K.~J.,  {James} G.~L.,  {Sykes} P.,  {Bird} T.~S.,  {Sinclair}
  M.~W.,  {Moorey} G.,    {Manchester} R.~N.,  2005, IEEE Antennas and
  Propagation Magazine, 47, 13

\bibitem[\protect\citeauthoryear{{Hou}, {Han} \& {Shi}}{{Hou}
  et~al.}{2009}]{hou09}
{Hou} L.~G.,  {Han} J.~L.,    {Shi} W.~B.,  2009, \aap, 499, 473

\bibitem[\protect\citeauthoryear{{Israel}, {Oosterbroek}, {Stella}, {Campana},
  {Mereghetti} \& {Parmar}}{{Israel} et~al.}{2001}]{israel01}
{Israel} G.,  {Oosterbroek} T.,  {Stella} L.,  {Campana} S.,  {Mereghetti} S.,
    {Parmar} A.~N.,  2001, \apjl, 560, L65

\bibitem[\protect\citeauthoryear{{Israel}, {Rea}, {Zelati}, {Esposito},
  {Burgay}, {Mereghetti}, {Possenti} \& {Tiengo}}{{Israel}
  et~al.}{2014}]{israel14}
{Israel} G.~L.,  {Rea} N.,  {Zelati} F.~C.,  {Esposito} P.,  {Burgay} M.,
  {Mereghetti} S.,  {Possenti} A.,    {Tiengo} A.,  2014, The Astronomer's
  Telegram, 6370, 1

\bibitem[\protect\citeauthoryear{{Kuiper}, {Hermsen}, {den Hartog} \&
  {Collmar}}{{Kuiper} et~al.}{2006}]{kuiper06}
{Kuiper} L.,  {Hermsen} W.,  {den Hartog} P.~R.,    {Collmar} W.,  2006, \apj,
  645, 556

\bibitem[\protect\citeauthoryear{{Paczynski}}{{Paczynski}}{1992}]{paczynski92}
{Paczynski} B.,  1992, Acta Astronomica, 42, 145

\bibitem[\protect\citeauthoryear{{Pavlovi{\'c}}, {Uro{\v s}evi{\'c}},
  {Vukoti{\'c}}, {Arbutina} \& {G{\"o}ker}}{{Pavlovi{\'c}}
  et~al.}{2013}]{pavlovic13}
{Pavlovi{\'c}} M.~Z.,  {Uro{\v s}evi{\'c}} D.,  {Vukoti{\'c}} B.,  {Arbutina}
  B.,    {G{\"o}ker} {\"U}.~D.,  2013, \apjs, 204, 4

\bibitem[\protect\citeauthoryear{{Perna} \& {Gotthelf}}{{Perna} \&
  {Gotthelf}}{2008}]{perna08}
{Perna} R.,  {Gotthelf} E.~V.,  2008, \apj, 681, 522

\bibitem[\protect\citeauthoryear{{Perna} \& {Pons}}{{Perna} \&
  {Pons}}{2011}]{perna11}
{Perna} R.,  {Pons} J.~A.,  2011, \apjl, 727, L51

\bibitem[\protect\citeauthoryear{{Pons} \& {Rea}}{{Pons} \&
  {Rea}}{2012}]{pons12}
{Pons} J.~A.,  {Rea} N.,  2012, \apjl, 750, L6

\bibitem[\protect\citeauthoryear{Rea \& Esposito}{Rea \&
  Esposito}{2011}]{rea11}
Rea N.,  Esposito P.,  2011, in Torres D.~F.,  Rea N.,  eds, Astrophysics and
  Space Science Proceedings, High-Energy Emission from Pulsars and their
  Systems.
Springer, Heidelberg, pp 247--273

\bibitem[\protect\citeauthoryear{{Rea}, {Esposito}, {Turolla}, {Israel},
  {Zane}, {Stella}, {Mereghetti}, {Tiengo}, {G{\"o}tz}, {G{\"o}{\u g}{\"u}{\c
  s}} \& {Kouveliotou}}{{Rea} et~al.}{2010}]{rea10}
{Rea} N.,  {Esposito} P.,  {Turolla} R.,  {Israel} G.~L.,  {Zane} S.,  {Stella}
  L.,  {Mereghetti} S.,  {Tiengo} A.,  {G{\"o}tz} D.,  {G{\"o}{\u g}{\"u}{\c
  s}} E.,    {Kouveliotou} C.,  2010, Science, 330, 944

\bibitem[\protect\citeauthoryear{{Rea}, {Israel}, {Pons}, {Turolla},
  {Vigan{\`o}}, {Zane}, {Esposito}, {Perna}, {Papitto}, {Terreran}, {Tiengo} \&
  {Salvetti}}{{Rea} et~al.}{2013}]{rea13}
{Rea} N.,  {Israel} G.~L.,  {Pons} J.~A.,  {Turolla} R.,  {Vigan{\`o}} D.,
  {Zane} S.,  {Esposito} P.,  {Perna} R.,  {Papitto} A.,  {Terreran} G.,
  {Tiengo} A.,    {Salvetti} D.,  2013, \apj, 770, 65

\bibitem[\protect\citeauthoryear{{Rea}, {Israel}, {Turolla}, {Esposito},
  {Mereghetti}, {G{\"o}tz}, {Zane}, {Tiengo}, {Hurley}, {Feroci}, {Still} \&
  {Yershov}}{{Rea} et~al.}{2009}]{rea09}
{Rea} N.,  {Israel} G.~L.,  {Turolla} R.,  {Esposito} P.,  {Mereghetti} S.,
  {G{\"o}tz} D.,  {Zane} S.,  {Tiengo} A.,  {Hurley} K.,  {Feroci} M.,  {Still}
  M.,    {Yershov} V.,  2009, \mnras, 396, 2419

\bibitem[\protect\citeauthoryear{{Rea}, {McLaughlin}, {Gaensler}, {Slane},
  {Stella}, {Reynolds}, {Burgay}, {Israel}, {Possenti} \& {Chatterjee}}{{Rea}
  et~al.}{2009}]{rea09b}
{Rea} N.,  {McLaughlin} M.~A.,  {Gaensler} B.~M.,  {Slane} P.~O.,  {Stella} L.,
   {Reynolds} S.~P.,  {Burgay} M.,  {Israel} G.~L.,  {Possenti} A.,
  {Chatterjee} S.,  2009, \apjl, 703, L41

\bibitem[\protect\citeauthoryear{{Rea}, {Oosterbroek}, {Zane}, {Turolla},
  {M{\'e}ndez}, {Israel}, {Stella} \& {Haberl}}{{Rea} et~al.}{2005}]{roz05}
{Rea} N.,  {Oosterbroek} T.,  {Zane} S.,  {Turolla} R.,  {M{\'e}ndez} M.,
  {Israel} G.~L.,  {Stella} L.,    {Haberl} F.,  2005, \mnras, 361, 710

\bibitem[\protect\citeauthoryear{{Rea}, {Pons}, {Torres} \& {Turolla}}{{Rea}
  et~al.}{2012}]{rea12}
{Rea} N.,  {Pons} J.~A.,  {Torres} D.~F.,    {Turolla} R.,  2012, \apjl, 748,
  L12

\bibitem[\protect\citeauthoryear{{Read}, {Rosen}, {Saxton} \& {Ramirez}}{{Read}
  et~al.}{2011}]{read11}
{Read} A.~M.,  {Rosen} S.~R.,  {Saxton} R.~D.,    {Ramirez} J.,  2011, \aap,
  534, A34

\bibitem[\protect\citeauthoryear{{Stamatikos}, {Malesani}, {Page} \&
  {Sakamoto}}{{Stamatikos} et~al.}{2014}]{stamatikos14}
{Stamatikos} M.,  {Malesani} D.,  {Page} K.~L.,    {Sakamoto} T.,  2014, GRB
  Coordinates Network, 16520, 1

\bibitem[\protect\citeauthoryear{{Staveley-Smith}, {Wilson}, {Bird}, {Disney},
  {Ekers}, {Freeman}, {Haynes}, {Sinclair}, {Vaile}, {Webster} \&
  {Wright}}{{Staveley-Smith} et~al.}{1996}]{staveley96}
{Staveley-Smith} L.,  {Wilson} W.~E.,  {Bird} T.~S.,  {Disney} M.~J.,  {Ekers}
  R.~D.,  {Freeman} K.~C.,  {Haynes} R.~F.,  {Sinclair} M.~W.,  {Vaile} R.~A.,
  {Webster} R.~L.,    {Wright} A.~E.,  1996, Publications of the Astronomical
  Society of Australia, 13, 243

\bibitem[\protect\citeauthoryear{{Stil}, {Taylor}, {Dickey}, {Kavars},
  {Martin}, {Rothwell}, {Boothroyd}, {Lockman} \& {McClure-Griffiths}}{{Stil}
  et~al.}{2006}]{stil06}
{Stil} J.~M.,  {Taylor} A.~R.,  {Dickey} J.~M.,  {Kavars} D.~W.,  {Martin}
  P.~G.,  {Rothwell} T.~A.,  {Boothroyd} A.~I.,  {Lockman} F.~J.,
  {McClure-Griffiths} N.~M.,  2006, \aj, 132, 1158

\bibitem[\protect\citeauthoryear{{Str{\"u}der}, {Briel}, {Dennerl}, {Hartmann},
  {Kendziorra}, {Meidinger}, {Pfeffermann}, {Reppin} \&
  {Aschenbach}}{{Str{\"u}der} et~al.}{2001}]{struder01}
{Str{\"u}der} L.,  {Briel} U.,  {Dennerl} K.,  {Hartmann} R.,  {Kendziorra} E.,
   {Meidinger} N.,  {Pfeffermann} E.,  {Reppin} C.,    {Aschenbach} B.,  2001,
  \aap, 365, L18

\bibitem[\protect\citeauthoryear{{Sun}, {Reich}, {Reich}, {Xiao}, {Gao} \&
  {Han}}{{Sun} et~al.}{2011}]{sun11}
{Sun} X.~H.,  {Reich} P.,  {Reich} W.,  {Xiao} L.,  {Gao} X.~Y.,    {Han}
  J.~L.,  2011, \aap, 536, A83

\bibitem[\protect\citeauthoryear{{Surnis}, {Krishnakumar}, {Maan}, {Joshi} \&
  {Manoharan}}{{Surnis} et~al.}{2014}]{pippone}
{Surnis} M.~P.,  {Krishnakumar} M.~A.,  {Maan} Y.,  {Joshi} B.~C.,
  {Manoharan} P.~K.,  2014, The Astronomer's Telegram, 6376, 1

\bibitem[\protect\citeauthoryear{{Turner}, {Abbey}, {Arnaud}, {Balasini},
  {Barbera}, {Belsole}, {Bennie}, {Bernard}, {Bignami} \& {Boer}}{{Turner}
  et~al.}{2001}]{turner01}
{Turner} M.~J.~L.,  {Abbey} A.,  {Arnaud} M.,  {Balasini} M.,  {Barbera} M.,
  {Belsole} E.,  {Bennie} P.~J.,  {Bernard} J.~P.,  {Bignami} G.~F.,    {Boer}
  M.,  2001, \aap, 365, L27

\bibitem[\protect\citeauthoryear{{Turolla}, {Zane}, {Pons}, {Esposito} \&
  {Rea}}{{Turolla} et~al.}{2011}]{turolla11}
{Turolla} R.,  {Zane} S.,  {Pons} J.~A.,  {Esposito} P.,    {Rea} N.,  2011,
  \apj, 740, 105

\bibitem[\protect\citeauthoryear{{Vigan{\`o}}, {Rea}, {Pons}, {Perna},
  {Aguilera} \& {Miralles}}{{Vigan{\`o}} et~al.}{2013}]{vigano13}
{Vigan{\`o}} D.,  {Rea} N.,  {Pons} J.~A.,  {Perna} R.,  {Aguilera} D.~N.,
  {Miralles} J.~A.,  2013, \mnras, 434, 123

\bibitem[\protect\citeauthoryear{{Weisskopf}, {Tananbaum}, {Van Speybroeck} \&
  {O'Dell}}{{Weisskopf} et~al.}{2000}]{weisskopf00}
{Weisskopf} M.~C.,  {Tananbaum} H.~D.,  {Van Speybroeck} L.~P.,    {O'Dell}
  S.~L.,  2000, in X-Ray Optics, Instruments, and Missions III. Edited by
  {Truemper}, J.~E. and {Aschenbach}, B. Proceedings of the SPIE. SPIE,
  Bellingham WA Vol.~4012 of Society of Photo-Optical Instrumentation Engineers
  (SPIE) Conference Series, {Chandra X-ray Observatory (CXO): overview}.
pp 2--16

\bibitem[\protect\citeauthoryear{{Wilms}, {Allen} \& {McCray}}{{Wilms}
  et~al.}{2000}]{wilms00}
{Wilms} J.,  {Allen} A.,    {McCray} R.,  2000, \apj, 542, 914

\bibitem[\protect\citeauthoryear{{Younes}, {Kouveliotou}, {Kargaltsev},
  {Pavlov}, {G{\"o}{\u g}{\"u}{\c s}} \& {Wachter}}{{Younes}
  et~al.}{2012}]{younes12}
{Younes} G.,  {Kouveliotou} C.,  {Kargaltsev} O.,  {Pavlov} G.~G.,  {G{\"o}{\u
  g}{\"u}{\c s}} E.,    {Wachter} S.,  2012, \apj, 757, 39

\end{thebibliography}

\bsp

\label{lastpage}

\end{document}